\documentclass[aps,prd,twocolumn,superscriptaddress,floatfix]{revtex4}
\usepackage{amsmath, amsfonts}
\usepackage{graphicx}
\newcommand{\beq}{\begin{equation}}
\newcommand{\eeq}{\end{equation}}
\newcommand{\beqn}{\begin{eqnarray}}
\newcommand{\eeqn}{\end{eqnarray}}
\newcommand{\rbl}{r_{\rm BL}}

\begin{document}
\title{Evolution of near-extremal-spin black holes using 
the moving puncture technique}
\date{\today}
\author{Yuk Tung Liu}
\affiliation{Department of Physics, University of Illinois at
  Urbana-Champaign, Urbana, IL 61801}
\author{Zachariah B. Etienne}
\affiliation{Department of Physics, University of Illinois at
  Urbana-Champaign, Urbana, IL 61801}
\author{Stuart~L.~Shapiro}
\altaffiliation{Also at Department of Astronomy and NCSA, University of
  Illinois at Urbana-Champaign, Urbana, IL 61801}
\affiliation{Department of Physics, University of Illinois at
  Urbana-Champaign, Urbana, IL 61801}

\begin{abstract}
We propose a new radial coordinate to write the Kerr metric in
puncture form. Unlike the quasi-radial coordinate introduced previously, 
the horizon radius remains finite in our radial coordinate in the 
extreme Kerr limit $a/M\rightarrow 1$. This significantly improves 
the accuracy of the evolution of black holes with spins close 
to the extreme Kerr limit. 
We are able to evolve 
accurately both stationary and boosted black holes with spins
as high as $a/M=0.99$ using initial data constructed in these new 
puncture coordinates. 
Initial data of compact binaries with rapidly spinning black holes 
can be constructed using our proposed new puncture metric for the 
background conformal metric. Our simulations for single black 
holes suggest that such initial data can be evolved successfully 
by the moving puncture technique.
\end{abstract}
\maketitle

\section{Introduction}

Binary black holes (BHBHs) are among the most promising sources 
of gravitational waves detectable by gravitation-wave detectors 
such as LIGO~\cite{LIGO1,LIGO2}, VIRGO~\cite{VIRGO1,VIRGO2}, 
GEO~\cite{GEO}, and TAMA~\cite{TAMA1,TAMA2}, as
well as by the proposed space-based interferometers LISA~\cite{LISA}, 
BBO~\cite{bbo-nasa} and DECIGO~\cite{DECIGO}. 
Supermassive black holes are likely formed during hierarchical mergers 
of halos and galaxies in the early universe. Binary black hole 
coalescence triggered by these mergers, followed by 
gas accretion onto the remnant hole, may give rise to 
a population of black holes with very rapid spin~\cite{vmqr05,bv08}. 
Standard thin-disk accretion alone spins up black holes (BHs) to 
a maximum value of $a/M=0.998$~\cite{thorne74}, 
where $a$ is the specific angular momentum and $M$ is the mass of 
the BH. However, accretion in thick, magnetized disks tends to drive the 
BH spin to $a/M \approx 0.94$~\cite{gsm04}. There is observational 
evidence suggesting that rapidly spinning BHs might exist in 
quasars~\cite{wchm06} and binary X-ray 
sources~\cite{mdgd06,mrfmg09,gmlnsrodes09}. 

There is great interest in studying rapidly spinning BHs 
in a compact binary system. The coalesence of rapidly spinning BHs 
could result in a gravitational-wave induced recoil velocity of a few 
thousand kilometers per second in some 
systems~\cite{clzm07a,clzm07b,ghsbh07,dlz08,hhhslm09}. Such recoil 
may have significant influence on the hierarchical evolution 
of supermassive BHs in galaxies~\cite{v07,vccdm08,bl08} and  
have observable signatures in quasars and active galactic nuclei~\cite{km08,gm08}. 
Black hole-neutron star binaries with a rapidly spinning BH may 
produce a substantial disk about the BH after 
merger~\cite{RKLRasio,elsb09,rj09}, 
which may be crucial to the formation of a short-hard gamma-ray burst.

Currently, the most common method of evolving compact binary systems in 
numerical relativity is the moving puncture 
technique~\cite{Campanelli:2005dd,Goddard1}. This technique requires 
initial data everywhere on the computational grid, including the BH 
interior. Most simulations 
adopt conformally-flat, puncture initial data. However, this 
type of initial data can only produce BHs with spins as high as 
$\approx 0.93$~\cite{dlt02,dlz08,lopc08,hhm09}, the extremal-Bowen-York 
limit. Moreover, conformally-flat initial 
data contain spurious gravitational waves even for isolated spinning BHs 
and thus cannot represent exact 
stationary Kerr BH spacetimes. BHBHs with BH spins close 
to this limit have been evolved using the moving puncture 
technique~\cite{mtbsg08,dlz08,hhm09}. Initial data with BH spins 
higher than 0.93 have been constructed using a (non-conformally flat) 
Kerr-Schild background metric~\cite{lopc08}, which for isolated BHs 
does not contain 
spurious radiation. These initial data have been 
evolved successfully using the generalized harmonic formalism with 
excision, even for BHs with spins higher than the extremal-Bowen-York
limit~\cite{lopc08}. One might wonder if these initial data 
can also be evolved by the moving puncture technique. Since these 
data are excised at the horizon, it is first necessary to fill in data 
everywhere inside the horizon in order to evolve the spacetime by 
the standard moving puncture technique. 

We have investigated the possibility of integrating Kerr-Schild 
initial data for a single, stationary, rotating BH using the 
moving puncture technique. We removed the physical 
ring singularity inside the horizon by filling the 
BH interior with constraint-violating 
``junk'' initial data. It has been demonstrated that 
the BSSN (Baumgarte-Shapiro-Shibata-Nakamura) 
scheme~\cite{SN,BS}, coupled with moving puncture gauge conditions, guarantees 
that the ``junk'' data will not propagate out of the 
horizon~\cite{eflsb07,bsstdhp07,bdsst09}. We have tried various methods of 
filling in the ``junk'' data, and are 
able to evolve the Kerr-Schild metric for a single BH with spins as high  
as $a/M=0.96$. However, when the BH spin exceeds 0.96, our code either crashes 
or the evolution becomes inaccurate (e.g.\ the BH's mass and spin deviate 
from their initial values significantly) after $\sim 10M$.

We next considered (nonconformally flat) puncture initial data that allow the BH spin to 
approach the Kerr limit. Brandt and Seidel have constructed such 
initial data~\cite{brandt-seidel95,brandt-seidel96}, which provide 
an exact description of Kerr spacetime with no spurious gravitational 
waves. Their metric generalizes 
the Schwarzschild metric in isotropic coordinates 
to rotating BHs. We are able to evolve this metric successfully 
using the moving puncture technique. 
However, when the BH spin approaches the extreme Kerr limit, the radius of
the BH horizon shrinks to zero in their quasi-isotropic radial coordinate. 
We find that this shrinkage causes numerical inaccuracy during the early 
evolution, which 
results in a slow decrease in the BH spin at late times 
(see Sec.~\ref{sec:results}).
In this paper, we introduce a new 
radial coordinate such that the horizon coordinate radius remains nonzero 
in the extreme Kerr limit. We are able to evolve accurately the puncture 
data in this new coordinate, both for stationary and boosted BHs with spins 
as high as $a/M=0.99$. 
Initial data for compact binaries with rapidly spinning BHs may be constructed
by using a conformal background metric consisting of
the superposition of two Kerr puncture 
metrics~\cite{dain01a,dain01b,hhbgs07,kp98}
in our proposed coordinates.
The simulations reported below suggest that such initial data can be 
evolved successfully by the moving puncture technique.

This paper is organized as follows: In Sec.~\ref{sec:formal}, we 
introduce our new puncture initial data, and briefly describe our 
numerical method to evolve the spacetime. We present results 
of our simulations in 
Sec.~\ref{sec:results}. We conclude in Sec.~\ref{sec:con} with a 
brief discussion of future applications of our technique.

\section{Formulation}
\label{sec:formal}

\subsection{Initial data}

We start from the Kerr metric in Boyer-Lindquist coordinates 
$(\rbl,\theta,\phi)$. We introduce the radial coordinate $\eta$ 
as in~\cite{brandt-seidel95,brandt-seidel96}:
\beq
  \rbl = r_+ \cosh^2 (\eta/2) - r_- \sinh^2 (\eta/2) \label{eq:eta1} \ , 
\eeq
where $M$ is the BH's mass, $a$ is the specific angular momentum, 
and $r_\pm = M\pm \sqrt{M^2-a^2}$ are the Boyer-Lindquist radii of
the outer (+) and inner ($-$) horizons of the BH. 
Both regions $\eta \in [0,\infty)$ 
and $\eta \in (-\infty,0]$ map to $\rbl \in [r_+,\infty)$. The BH 
event horizon $\rbl=r_+$ is mapped to $\eta=0$. Equation~(\ref{eq:eta1}) is 
invariant under the inversion $\eta \rightarrow -\eta$. The spatial metric 
in this coordinate system is given by 
\beq
  \gamma_{ij}dx^i dx^j = \Psi_0^4 \left[ e^{-2q_0} (d\eta^2+d\theta^2)
  + \sin^2 \theta d\phi^2 \right] \ ,
\label{eq:gammaij}
\eeq
where $\Psi_0^4 =A/\Sigma$, $e^{-2q_0} = \Sigma^2/A$, 
$\Sigma = \rbl^2 + a^2 \cos^2 \theta$, 
$\Delta = \rbl^2-2M\rbl +a^2$, and $A=(\rbl^2+a^2)^2 -\Delta a^2\sin^2 \theta$.
%%\beqn
%%  \Psi_0^4 &=& A/\Sigma \ \ , \ \  e^{-2q_0} = \Sigma^2/A \ , \\
%%  \Sigma &=& \rbl^2 + a^2 \cos^2 \theta  \ \ , \ \ 
%%\Delta = \rbl^2-2M\rbl +a^2 \\
%%  A &=& (\rbl^2+a^2)^2 -\Delta a^2\sin^2 \theta \ .
%%\eeqn
The spatial metric is invariant under the inversion $\eta \rightarrow -\eta$, 
and is asymptotically flat at $\eta \rightarrow \pm \infty$. 
The BH exterior is mapped twice in this metric and the two pieces 
are joined smoothly at the throat $\eta=0$. The metric describes an 
Einstein-Rosen bridge. The nonzero components of the extrinsic curvature are 
\beqn
  K_{ij} &=& \Psi_0^{-2} \hat{K}_{ij} \\
  \hat{K}_{\eta \phi} &=& \hat{K}_{\phi \eta} = \frac{Ma \sin^2 \theta}
{\Sigma^2}\times \cr
  && [2\rbl^2 (\rbl^2+a^2)+\Sigma (\rbl^2-a^2)] \\
  \hat{K}_{\theta \phi} &=& \hat{K}_{\phi \theta} = -2Ma^3 \rbl 
\sqrt{\Delta} \cos \theta \sin^3\theta /\Sigma^2 \ .
\eeqn
The lapse and shift that give rise to a stationary 
spacetime are 
\beqn
  \alpha &=& \sqrt{\frac{\Delta \Sigma}{A}} \label{kerr_lapse} \\
  \beta^\phi &=& -\frac{2Ma\rbl}{A} \ \ \ , \ \ \
  \beta^\eta = \beta^\theta=0 \ . \label{kerr_shift}
\eeqn
%\beqn   
%  \alpha &=& \sqrt{\Delta \Sigma/A} \label{kerr_lapse} \\
%  \beta^\phi &=& -2Ma\rbl/A \ \ \ , \ \ \ 
%  \beta^\eta = \beta^\theta=0 \ . \label{kerr_shift}
%\eeqn

Brandt and Seidel introduce a quasi-isotropic radial 
coordinate~\cite{brandt-seidel95,brandt-seidel96}: 
\beq
  \bar{r} = \frac{\sqrt{M^2-a^2}}{2}e^\eta \ .
\label{eq:riso1}
\eeq
It follows from Eq.~(\ref{eq:eta1}) that 
\beq
  \rbl = \bar{r} \left( 1+\frac{M+a}{2\bar{r}}\right)
\left( 1+\frac{M-a}{2\bar{r}}\right) \ . 
\eeq
In the Schwarzschild limit $a=0$, the spatial metric reduces to 
the Schwarzschild metric in isotropic coordinates. 
This quasi-isotropic radial coordinate 
has an undesirable property that the BH horizon at $\eta=0$ corresponds to 
$\bar{r}=\sqrt{M^2-a^2}/2$, which shrinks to zero in the extreme Kerr limit. 
To reduce this numerical inconvenience, we generalize Eq.~(\ref{eq:riso1}) 
by considering a radial coordinate of the form 
\beq
  r= \frac{\sqrt{M^2-a^2}}{2} \lambda(a,\eta) e^\eta \ ,
\label{eq:risoGen}
\eeq
where $\lambda(a,\eta)$ is an arbitrary function of $a$ and $\eta$. 
One seeks to choose $\lambda$ such that (1) $\eta=0$ corresponds to a 
nonzero value of $r$ for any value of $|a| \leq M$, 
(2) $\lambda=1$ when $a=0$ and (3) $\lambda \rightarrow 1$ as 
$\eta \rightarrow \pm \infty$. Property (2) ensures that the 
usual isotropic radial coordinate is recovered in the Schwarzschild limit. 
Property (3) ensures that $r \rightarrow \rbl$ at spatial infinity. 
One simple choice of $r$ that satisfies all three properties is 
given by 
\beq
  \rbl = r \left( 1 + \frac{r_+}{4r} \right)^2 \ ,
\label{eq:risokerr}
\eeq
which corresponds to setting $\lambda$ according to
\beq
  \lambda = \frac{e^{-\eta}}{\sqrt{M^2-a^2}} \left[ 
\rbl - \frac{r_+}{2} + \sqrt{\rbl (r_+-r_-)} \sinh(\eta/2)\right] .
\eeq
%%\beqn
%%  & \frac{1}{\lambda} = \frac{1}{2r^2} \left[ \left(r+\frac{r_+}{4}\right)^2
%% - Mr \right. & \cr && \cr   & \left. + \sqrt{ \left(r + \frac{r_+}{4}\right)^4 - 2Mr \left(r + \frac{r_+}{4}\right)^2
%%+ a^2 r^2} \right] \ . &
%%\eeqn
The regions $\eta \in (-\infty,0]$ and $\eta \in [0,\infty)$ are mapped to
$r\in (0,r_+/4]$ and $r \in [r_+/4,\infty)$, respectively.
The horizon is located at $r=r_+/4$. In the extreme Kerr limit, the 
horizon radius is $r=M/4>0$. The spatial metric and extrinsic curvature 
in this new coordinate system are given by 
\beqn
  {}^{(3)}ds^2 &=& \frac{\Sigma \left(r+\frac{r_+}{4}\right)^2}
{r^3 (\rbl-r_-)}dr^2 + \Sigma d\theta^2
+\frac{A\sin^2 \theta}{\Sigma}d\phi^2 \label{eq:Kerr_puncture} \\
  K_{r\phi} &=& K_{\phi r} = \frac{Ma\sin^2\theta}{\Sigma \sqrt{A \Sigma}}
[ 3\rbl^4 + 2a^2\rbl^2-a^4 \cr
&& -a^2(\rbl^2-a^2)\sin^2\theta ] \left( 1 +\frac{r_+}{4r}\right) \times \cr && \cr && \frac{1}{\sqrt{r(\rbl-r_-)}} \ , \\   K_{\theta \phi} &=& K_{\phi \theta} = -\frac{2a^3M\rbl \cos\theta \sin^3
\theta}{\Sigma \sqrt{A\Sigma}} \left(r-\frac{r_+}{4}\right) \times \cr
&& \sqrt{\frac{\rbl-r_-}{r}} \ . 
\eeqn

In our numerical evolution, we use Cartesian coordinates $(x,y,z)$, 
which are related to the $(r,\theta,\phi)$ coordinates by the usual 
transformation: $x=r \sin \theta \cos \phi$, $y=r \sin \theta \sin \phi$ 
and $z=r \cos \theta$. Cartesian components of the spatial metric 
$\gamma_{ij}$ and extrinsic curvature $K_{ij}$ are 
computed by the usual transformation formula 
of tensor components. The initial data of a rotating BH moving with 
a constant velocity as measured by a distant observer are constructed by boosting the spacetime metric 
derived from Eq.~(\ref{eq:Kerr_puncture}) and the lapse and shift 
in Eqs.~(\ref{kerr_lapse}) and (\ref{kerr_shift}). 

\subsection{Numerical evolution scheme}

The formulation and numerical scheme for our simulations are basically
the same as those already reported in~\cite{eflstb08,elsb09}, to which
the reader may refer for details. We adopt the 
BSSN formalism coupled to the standard moving puncture gauge conditions 
to evolve the spatial metric and extrinsic curvature. The 
evolution equations are given by Eqs.~(9)--(13) in~\cite{eflstb08}. 
The gauge conditions are given by Eqs.~(2)--(4) in~\cite{elsb09}, 
with the gauge parameter $\eta$ set to $1/M$. 
During the evolution, we adopt Eqs.~(29), (30) in~\cite{eflstb08} and 
Eq.~(11) in~\cite{dsy04} to help enforce/control additional constraints 
in the BSSN variables.

We evolve the BSSN equations
with sixth-order accurate, centered finite-differencing stencils,
except on shift advection terms, where we use sixth-order accurate
upwind stencils.  We apply Sommerfeld outgoing wave boundary
conditions to all BSSN fields. Our code is embedded in
the Cactus parallelization framework~\cite{Cactus}, and our
fourth-order Runge-Kutta timestepping is managed by the {\tt MoL}
(Method of Lines) thorn, with a Courant-Friedrichs-Lewy (CFL) factor
set to 0.25 in all simulations.  We find that we get better results 
if we add a seventh-order Kreiss-Oliger dissipation of the form 
\beq  
(\epsilon/256) (\Delta x^7 \partial_x^7 + \Delta y^7 \partial_y^7 
+ \Delta z^7 \partial_z^7) f
\eeq
to the lapse and shift variables $f$, with the parameter $\epsilon$ set 
to 0.9. We use the
Carpet~\cite{Carpet} infrastructure to implement moving-box
adaptive mesh refinement. The apparent horizon of the
BH is computed with the {\tt AHFinderDirect} Cactus
thorn~\cite{ahfinderdirect}. The BH's mass and angular momentum 
are computed using the isolated horizon formalism~\cite{ak04}, 
with the axial Killing vector computed using the numerical 
technique described in~\cite{dkss03}. 

For the initial lapse and shift, we have implemented the lapse and 
shift obtained from the analytic spacetime metric (\ref{kerr_lapse}), 
(\ref{kerr_shift}) and (\ref{eq:Kerr_puncture}) 
(boosted in the case of a moving BH),
as well as the standard choice setting $\alpha=\psi^{-2}$ and
$\beta^i=0$ [where $\psi=(\mbox{det$\gamma_{ij}$)}^{1/3}$]. We find that
these two different sets of initial lapse and shift data yield 
similar evolution results for stationary BHs. The first set of 
lapse and shift yields a slightly better result for boosted BHs.
We show the results for the second set of initial lapse and shift 
for stationary BHs and the first set for boost BHs in the next section. 

\section{Results}
\label{sec:results}

\begin{figure}
\includegraphics[width=9cm]{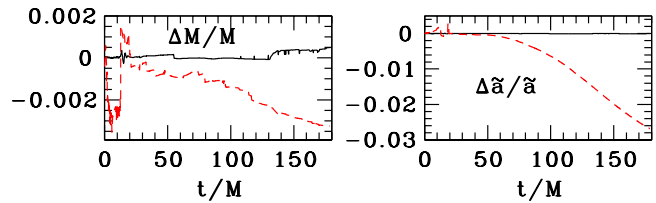}
\caption{Fractional error of the BH mass $\Delta M/M$ (left graph) 
and spin parameter $\Delta \tilde{a}/\tilde{a}$ (right graph) 
vs time for a stationary BH with spin parameter $\tilde{a}\equiv a/M=0.99$. 
Dash (red) lines show the results for the quasi-isotropic radial 
coordinate, and solid (black) lines show the results for our 
new radial coordinate. 
The resolution in the innermost refinement level is $M/50$ for both 
cases.}
\label{fig:stBH}
\end{figure}

\begin{figure}
\includegraphics[width=6cm]{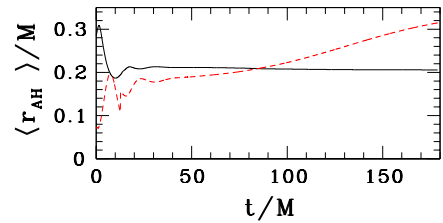}
\caption{Evolution of the average coordinate radius of the BH's horizon evolved with  
our radial coordinate (black solid line) and the quasi-isotropic coordinate 
(red dashed line).}
\label{fig:bh_radius}
\end{figure}

\begin{figure}
\includegraphics[width=6cm]{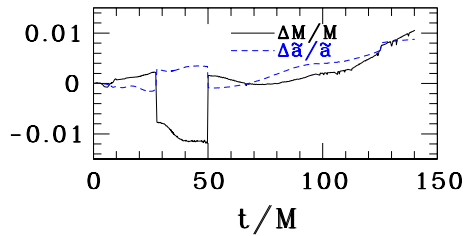}
\caption{Fractional error of the BH mass $\Delta M/M$ 
(black solid line)
and spin parameter $\Delta \tilde{a}/\tilde{a}$ (blue dash line)
vs time for a BH with spin parameter $\tilde{a}\equiv a/M=0.99$ and moving with 
a momentum $P=0.5M$ relative to observers at spatial infinity. The resolution 
in the innermost refinement level is $M/50$.}
\label{fig:movBH}
\end{figure}

We perform simulations on rapidly rotating BHs with spin 
parameter $\tilde{a}\equiv a/M=0.99$ for cases where 
the BH is stationary and boosted to give 
a momentum $P=0.5M$, relative to 
observers at spatial infinity. We use seven refinement levels 
for all these simulations. The resolution in the innermost refinement level 
is $M/50$ for a typical run. We also try resolutions $M/40$,
$M/60$ and $M/80$ for the stationary BH case and confirm that 
the code converges at approximately fourth order, as expected. 
The outer boundary is 
a rectangular box with a length of $512M\times 512M\times 256M$ in the 
$(x,y,z)$ directions. 
We impose reflection symmetry about the 
equatorial ($z=0$) plane. 

Figure~\ref{fig:stBH} shows the fractional error in the 
mass and spin parameter of the BH as a function of time for a 
stationary BH, evolved with both the quasi-isotropic radial coordinate 
and our proposed new radial coordinate. The grid structure, resolution and 
gauge conditions are identical for these two runs. We see that the BH's spin slowly 
decreases with time when evolved with the quasi-isotropic radial coordinate. 
Such a secular shift of the BH's spin is also observed in the evolution of 
a near-extremal-Bowen-York-spin BH, and can be reduced by using higher 
resolution~\cite{mtbsg08}. By contrast, we see no such drift when evolved 
with our proposed radial coordinate using the same resolution. 
The BH's spin is conserved to within 
$10^{-4}$ during the entire evolution of $180M$. We attribute this result 
to the fact that the BH interior is better resolved with our radial 
coordinate during the early simulation. At $t=0$, the coordinate radius 
of the horizon is $0.07M$ in the quasi-isotropic coordinate and $0.285M$ in 
our radial coordinate. Hence the initial BH's diameter is covered by 
7 grid points in the quasi-isotropic coordinate and 28 grid points in our 
radial coordinate. Figure~\ref{fig:bh_radius} shows the average coordinate 
radius of the BH as a function of time, evolved with our radial coordinate 
(black solid line) and the quasi-isotropic coordinate (red dashed 
line). We find that the average coordinate radius approaches a constant 
value after $t \gtrsim 50M$ when evolved with our radial coordinate, and
the metric approaches a ``trumpet'' geometry~\cite{hhpbm07,hhobm08} 
in which the conformal factor $\psi \propto r^{-1/2}$ near the puncture.
By contrast,  
we find the radius increases slowly at late time when evolved with the 
quasi-isotropic coordinates, which correlates with the slow decrease 
in the BH spin due to accumulated numerical inaccuracy during the 
early evolution.

We have tried to evolve a BH with spin parameter $a/M=0.999$. We find that 
although the initial horizon radius is $0.261M$, the puncture evolution 
quickly drives the horizon radius to below $0.1M$ after $\sim 5M$. The 
BH spin slowly decreases with time due to insufficient resolution. 
It has been reported that the horizon radius increases 
when the parameter $\eta$ in the shift equation 
increases~\cite{bghhst08}. 
We have observed this behavior 
for lower spin BHs. However, we find that the evolution of the horizon 
size is insensitive to the values of $\eta$ for the high-spin BHs. For 
example, for a BH with $a/M=0.99$, the horizon sizes evolved with different 
values of $\eta$ change by less than 5\%.
Our numerical experiments seem to suggest that the puncture evolution
will eventually drive the coordinate size of the horizon to a constant 
value tending towards zero as the BH
spin approaches the extreme Kerr limit. However, for a given BH spin 
$a$ close to the extreme Kerr limit, the final size of the 
horizon is still larger than the initial horizon radius
in the quasi-isotropic coordinate. Hence our proposed radial coordinate
is better suited for evolving high-spin BHs than the quasi-isotropic
coordinate.

Finally, Fig.~\ref{fig:movBH} shows the boosted case in which 
the BH moves with a momentum $P=0.5M$. 
We see that the errors are less than 1.2\% throughout the
entire evolution lasting about $150M$, during which the BH has traveled a
coordinate distance of about $60M$. This demonstrates that stable 
evolution of rapidly rotating BHs can be achieved using the moving 
puncture technique for the puncture initial data in our radial coordinate.

\section{Conclusion}
\label{sec:con}

We construct a new radial coordinate to write the Kerr metric in 
puncture form. This new radial coordinate has the advantage 
that the BH horizon radius remains finite 
in the extreme Kerr limit, which is useful for numerical 
simulations. By contrast, the horizon radius approaches 
zero in the quasi-isotropic coordinate originally adopted 
for this metric. We have demonstrated that higher accuracy is achieved 
by using our coordinate rather than the quasi-isotropic coordinate when 
evolving a high-spin BH. With our new 
coordinate, we are able to evolve, 
using the moving puncture technique, rapidly rotating BHs, 
both stationary and boosted, with spin parameters as high as 0.99. 

Binary black hole initial data with rapidly spinning BHs may be 
constructed using a conformal background metric consisting of 
the superposition 
of two Kerr-like conformal metrics in puncture form. This type of 
initial data for binary black hole head-on collision 
has been constructed in quasi-isotropic 
coordinates~\cite{dain01a,dain01b,hhbgs07,kp98}. It will be useful 
to generalize this technique to construct 
quasi-circular, rapidly spinning binary black hole initial data 
using superposed puncture Kerr metrics 
in our proposed radial coordinate for the background metric. 
Our numerical results presented 
in this paper suggest that such initial data can be evolved 
reliably using the moving puncture technique. 
This type of initial data has the additional advantages  
that the BH spins can be higher and the amount of spurious 
gravitational radiation will be  
significantly less than the conformally flat initial data, 
as demonstrated in~\cite{hhbgs07} for the head-on collision 
case.

{\it Acknowledgments}: This paper was supported in part by NSF
Grants PHY02-05155 and PHY06-50377 as well as NASA
Grants NNG04GK54G and NNX07AG96G to the University of Illinois at
Urbana-Champaign.

\bibliography{paper.bib}
\end{document}